 \documentclass[twocolumn]{aa} 
\usepackage{graphicx}
\usepackage{txfonts}
\newcommand\vv{{\mathrm v}  }       

\begin{document}
   \title{Tidal effects on the radial velocity curve of HD77581  (Vela X-1)}

   \author{G. Koenigsberger
       \inst{1}
   E. Moreno         
       \inst{2}
   \and
   D. M. Harrington
       \inst{3}
        }

   \offprints{G. Koenigsberger}

\institute{Instituto de Ciencias F\'{\i}sicas, Univer\-sidad Nacio\-nal
           Aut\'o\-noma de M\'exico, Cuernavaca, Mor. 62210, Mexico\\
           \email{gloria@astro.unam.mx}
\and
           Instituto de  Astrono\'{\i}a, Univer\-sidad Nacio\-nal Aut\'o\-noma de M\'exico, 04510, Mexico\\
           \email{edmundo@astro.unam.mx}
      \and
           Institute for Astronomy, University of Hawaii, {2680 Woodlawn Drive}, Honolulu, HI, 96822\\
           \email{dmh@ifa.hawaii.edu}
          } 
 \date{Received; accepted }

\abstract
{The mass of the neutron star in Vela X-1 has been found to be more massive than the canonical 1.5 M$_\odot$.
This result relies on the assumption that the amplitude of the optical component's measured radial velocity curve
is not seriously affected by the interactions  in the system. 
}
{Our aim is  to explore the effect on the radial velocity curve caused by surface motions excited by tidal 
interactions.
}
{We use a calculation from first principles that involves solving the equations of motion of a Lagrangian grid of
surface elements.  The velocities on the visible surface of the star 
are  projected along the line-of-sight to the observer to obtain the Doppler shifts which are applied to the 
local line-profiles, which are then combined to obtain the absorption-line profile in the observer's reference frame.
The centroid of the line-profiles for different orbital phases is then measured and a simulated  RV curve constructed. 
 Models are run for the ``standard" ($\vv~sini$=116 km/s) and ``slow" (56 km/s) supergiant
rotation velocities.
 }  
{The surface velocity field   is complex and includes fast, small-spatial scale structures.  It  leads to 
strong variability in the photospheric line profiles which, in turn, causes significant deviations from  
a Keplerian RV curve.   The peak-to-peak amplitudes of model RV curves  are in all cases larger
than the amplitude of the orbital motion.  Keplerian fits to RV curves obtained with the ``standard"        
rotation velocity imply $m_{ns}\geq$1.7 M$_\odot$. However, a similar analysis of the ``slow" rotational
velocity models allows for $m_{ns}\sim$1.5 M$_\odot$.  Thus, the stellar rotation plays an important role
in determining the characteristics of the perturbed RV curve.
}
{
Given the observational uncertainty in  GP Vel's projected rotation velocity and  the strong perturbations seen in 
the published and the model RV curves, we are unable to rule out a small ($\sim$1.5 M$_\odot$) 
mass for the neutron star companion.
}

  \authorrunning{Koenigsberger, Moreno  \& Harrington}
  \titlerunning{RV curve of Vela X-1}
   \maketitle

\section{Introduction}

The high mass X-ray binary system  Vela X-1 (4U 0900-40) consists of a
pulsar (P$_{pulse}$=283 s) in an eccentric (e=0.09, P$_{orb}$=8.96 d) orbit 
around the B0.5-supergiant star HD 77581=GP Vel. Particular interest in this system 
is sparked by the  mass derived for the  neutron star, M$_{ns}\sim$1.8 M$_\odot$, since 
this is significantly larger than the canonical 1.5 M$_\odot$ upper limit that is 
predicted by the standard neutron star equation of state  as well as that 
which is predicted from supernova models for the newly formed neutron stars.

The maximum possible mass of a neutron star depends on the equation of state of the
ultradense matter in its interior (Lattimer \& Prakash 2007; Page \& Reddy 2006). For 
nuclear densities $\sim$3$\times$10$^{14}$ g cm$^{-3}$, the equation of state is a 
``soft" one, and the maximum mass is $\sim$1.5 M$_\odot$.  For larger densities, 
the equation of state is a ``stiff" one, and depending on the adopted equation of state,  
the upper mass limit may be  as high as 2.9 M$_\odot$ (Kalogera \& Baym 1996).    The  
majority of neutron stars in binary systems do indeed have $m_{ns}\leq$1.5 M$_\odot$ 
(Thorsett \& Chakrabarty 1999; Schwab et al. 2010; Lattimer \& Prakash 2010). On the 
other hand, the existence of  $m_{ns}\sim$2 M$_\odot$ neutron stars  is now fairly well established 
(Demorest et al. 2010),  although it is not clear whether such high-mass neutron stars are created at
birth in the supernova event or are the consequence of subsequent accretion processes (Woosley \& Heger 
2007; Timmes et al. 1996). Van den Heuvel (2004) has argued that the GP Vel system provides direct 
evidence that there is at least one group of neutron stars that are indeed born with such a  large mass. 

The procedure for determining the  neutron star mass in binary systems such as GP Vel  is based
on the radial velocity (RV) curve which is derived from photospheric absorption lines observed in 
the spectrum of the optical companion.  Implied in this approach is the assumption 
that the RV variations truly represent the orbital motion of the star.   In the case of 
GP Vel, the validity of this assumption is highly questionable because
it has long been known that  the photospheric absorptions of GP Vel undergo
prominent line profile variability.  Specifically, the lines develop asymmetries
over the orbital cycle which lead to systematic deviations of the measured RV data points 
from a Keplerian RV curve. It is believed that the line-profile variability is associated 
with tidal forcing, non-radial  pulsations and other interaction effects in the binary system 
(van Paradijs et al. 1977b; van Kerkwijk et al. 1995; Barziv et al. 2001, Quaintrell et al. 2003). This 
raises the question of whether these systematic deviations may not lead to an artificially large RV curve
amplitude and, hence, an overestimate of the neutron star mass.

The tidal effects are clearly present, as shown by  the optical light curve which 
displays ellipsoidal variations with a full amplitude $\sim$0.1 mag  indicating that the star 
is strongly distorted by the neutron star's gravitational field  (Jones \& Liller 1973;
Zuiderwijk et al. 1977;  Tjemkes et al. 1986).
Van Paradijs et al. (1977b) explored the effects of the deformation of the star on the RV 
curves. They assumed a circular orbit and synchronous rotation and computed the shape of
photospheric absorption lines assuming a non-uniform temperature distribution.  However,
the rotation of GP Vel is clearly slower than synchronous, and it cannot be assumed to have a simple
Roche geometry.  Indeed, Tjemkes et al. (1986) showed that  a model for the
ellipsoidal variations based on the assumption of an ``equilibrium tide" does not adequately
reproduce the shape of the observed light curve.  They suggested that 
non-linear effects might be important,  but full non-linear calculations for
tidally interacting stars are only now becoming available (Weinberg et al. 2011).

We have developed a 2D calculation, the TIDES\footnote{Tidal Interactions with Dissipation of
Energy through Shear} code, that provides the time-dependent  shape of the stellar surface and
its  surface velocity field for the general case of an elliptic orbit and asynchronous rotation.
Using the derived velocity field,  the line-profile variability is computed (Moreno \& 
Koenigsberger 1999; Moreno, Koenigsberger \& Toledano 2005).  The method, though limited to 
the analysis of the surface layer, provides insight into the non-linear effects that appear
in binary stars  that are not in an equilibrium configuration.  In particular, it allows
the analysis of highly eccentric and asynchronous systems. Here, tidal flows are  present    
in forcing regimes that are far from equilibrium configurations,  and
linear approximations are inapplicable.  In this paper we use this model 
to explore the effects on the absorption line-profiles  produced by the tidal flows on the 
B-supergiant's surface, and the resulting deformation of the RV curve.

In Section of 2 of this paper we summarize the method for calculating the tidally-perturbed  RV curves;
Section 3 describes the method for chosing the stellar and binary parameters; Section 4 contains the
results; and Section 5 lists the conclusions.


\section{Calculation of tidally-perturbed RV curves}

\subsection{The TIDES code calculations}

Our method consists of computing  the motion of a Lagrangian grid of surface elements distributed along
a series of co-latitudes covering the  surface of the star with
mass $m_1$ as it is perturbed by its companion of mass $m_2$, which is assumed to be a point source.
The main stellar body below the perturbed layer is assumed to have uniform rotation. The equations of
motion that are solved for the set of surface elements include the gravitational fields of $m_1$ and $m_2$, the
Coriolis force, the centrifugal force, and gas pressure. The motions of all surface elements  are 
coupled through the  viscous stresses included in the equations of motion. The surface layer is 
coupled to the interior body of the star also through the viscosity. The simultaneous solution of the 
equations of motion for all surface elements yields values of the radial and azimuthal velocity fields 
over the stellar surface, $\vv_r$ and $\vv_\varphi$, respectively.

The calculated velocity field is then projected along the line-of-sight to the observer
to obtain the Doppler shifts required to produce the integrated photospheric absorption line 
profiles.  We assume a Gaussian shape for the local profiles\footnote{A  discussion
on the effect on the results of the intrinsic line-profile shape is given in Harrington et al. (2009); the
use of Gaussians is supported by the results of Landstreet et al. (2009).}
 and  a limb-darkening law of the form $s(\theta)=(1-u+u cos \theta)$, with $u$=0.6 as in the 
Milne-Eddington approximation.
Full details of the model are given in Moreno \& Koenigsberger (1999), Toledano et al. (2007) and  
Moreno et al. (2005, 2011). An application of the model to the B-type binary $\alpha$ Vir (Spica) 
may be found in Harrington et al. (2009).  

It is worth noting that the TIDES model  computes the surface motions of the optical component from 
first principles and allowing for non-linear effects.   However, there are currently two simplifications 
in the method. The first simplification is that the equations of motion are solved
only for a thin surface layer, instead of for the entire star,  which  has the disadvantage 
that the perturbations  from deeper layers in the star are neglected.  This includes the excitation of 
non-radial pulsation modes which have been shown to cause oscillations in the RV curve (Willems \& Aerts 2002).  
On the other hand, its response is  representative of the effects that are
produced on the outer stellar layer, which is the one that is most strongly affected by the
tidal forces (see, for example, Dolginov \& Smel'chakova 1992).  The second simplification is  that motions 
in the polar direction are suppressed, allowing only motions in the radial and azimuthal directions.
In addition,  we neglect the effects of  possible temperature variations across the stellar surface.  

The benefits of the model are: 1) we  make no {\it a priori} assumption regarding the mathematical
formulation of the tidal flow structure since we derive the velocity field  from
first principles; 2) the method is not limited to slow stellar rotation rates nor to small orbital
eccentricities; and 3) it is computationally inexpensive.

The parameters that are needed for the calculation are:  the orbital period, $P$, eccentricity, $e$, 
inclination, $i$, and argument of periastron, $\omega_{per}$; the stellar masses, $m_1$ and $m_2$, 
the radius of the primary star, $R_1$, its equatorial rotation speed, $\vv_{rot}$, the polytropic
index\footnote{The polytropic index of a {\bf radiative} supergiant star is $n\geq$3, Schwarzschild, 1958, p.258.}, 
$n$, and kinematical viscosity of its material, $\nu$.  In addition, the code requires 
the depth of the surface layer, $dR/R_1$ and the number surface elements for which the equations
of motion are to be solved.  The latter is specified by the number of longitudes into which the 
equator is divided and the number of latitudes between the equator and the polar region.  We have
done a thorough investigation of the TIDES code behavior (Harrington et al., in preparation)  and 
find that  500 partitions in the azimuthal direction at the equator and 20 latitudes is sufficient 
to resolve the small-scale structure.  This implies approximately 6800 surface elements\footnote{The 
number of azimuthal partitions decreases from the equatorial latitude to the polar region.} in the 
semi-hemisphere above and including the equator. Since the axis of stellar rotation is assumed to be
parallel to that of the orbital motion, the perturbations on the northern and southern hemispheres
are assumed to be symmetric.  Table 1 gives a description of the input parameters
and lists the values for those parameters which were held constant throughout all the calculations
in this paper.

The line-profile calculation is performed  after the initial transitory phase of the calculation
has damped down.  In the case of GP Vel, the steady state is attained after $\sim$20 orbital cycles, and
the line-profile calculation is performed at 30 orbital cycles.

\begin{table}
\caption{Description of input parameters for the TIDES code.\label{tbl-1}}
\label{table1}
\centering
\begin{tabular}{lllll}
\hline\hline
Parameter                 &         Description         & Values used           \\
\hline
M$_1$ (M$_\odot$)        &   mass of star 1                       &  see Table 3 \\
M$_2$ (M$_\odot$)        &   mass of star 2                       &  see Table 3 \\
R$_1$ (R$_\odot$)        & stellar radius at equilibrium          &  see Table 3  \\
P$_{orb}$ (days)         &  orbital period                        & 8.964368 d    \\
$e$                      &  orbital eccentricity                  & 0.0898        \\
$i$   (deg)              & inclination of  orbital plane          & 78.8, 85.9     \\
$\omega_{per}$ (deg)     & argument of periastron                 & 332             \\
$\beta_{0}$              & asynchronicity parameter               & see Table 3  \\
$\nu$ (R$^2$$_\odot$/day)& kinematical viscosity                  & 0.12, 0.22               \\
a                        & absorption-line depth                  & 0.7          \\
n                        & polytropic index                       & 3.0       \\
k                        & line-profile broadening param          & 30           \\
N$_{az}$                 & num.  segments along  equator          & 500                      \\
N$_{lat}$                & number of latitudes                    & 20                    \\
\hline
\hline
\end{tabular}
\end{table}

\subsection{RV measurements}

For each given set of input parameters,  the output of the  TIDES code yields two line-profile files, one 
containing  the line-profiles arising in the tidally perturbed surface and the second  containing 
the profiles arising from the unperturbed,  rigidly-rotating surface.  The  lines  are  measured 
to obtain their centroid.  The centroid was computed  through the weighted sum over all wavelengths,  
$\lambda_i$,  as follows:

\begin{equation}
\lambda_{center}=\frac{\Sigma \lambda_i  (C_i-I_i)^{3/2}}{\Sigma (C_i-I_i)^{3/2}}
\end{equation}

\noindent where the summation is carried out between the two limits within which the absorption
line lies.  In practice, the limits are the points where the absorption meets the continuum level.
$I_i$ and $C_i$ are the  line and continuum intensities, respectively.
This definition of the centroid is the same as that used by the 'e' function in the IRAF\footnote{{\it Image
Reduction and Analysis Facility}, distributed by NOAO} subroutine {\it splot}. 

The RVs were also measured through a Gaussian fit to the line profiles, using the {\it IDL} routine
GAUSSFIT. Although a Gaussian fit is a very poor approximation to the actual shape of the lines, the
derived RV's were in general very similar to those obtained with the flux-weighted centroid method. In
the remainder of this paper we use the flux-weighted centroids.

\subsection{Semi-amplitude of the RV curve}

The radial velocities that were obtained from the model line profiles were used to construct
the RV curves.  These curves were fit with the function that characterizes the orbital motion;
i.e.,  the function describing the  Keplerian orbit\footnote{Binnendijk, 1960,p.149--151}:
\begin{equation}
V_r=V_0 + K_1[e~cos~\omega_{per} + cos(\theta+\omega_{per})]
\end{equation}
\noindent where
\begin{equation}
K_1= \frac{2 \pi}{P} \frac{a_1~sin~i}{(1-e^2)^{1/2}}
\end{equation}
\noindent is the semi-amplitude of the RV curve, $a_1$ is the semi-major axis of
$m_1$'s orbit,  $V_0$ is a constant vertical offset (the ``gamma" velocity), and $\theta$ is
the true anomaly.

The fit was performed using the {\it Powell} minimization method in an {\it IDL} script\footnote{The POWELL algorithm as 
described in Press et al. 1992, Section 10.5}.  The free parameters for the fit were $a_1~sin~i$ and $V_0$.
The fitted value of $K_1$ was then computed using eq. (3).

\subsection{A note on orbital phases}

In order to be able to compare the theoretical models with the observational data, the
conventions for setting orbital phase $\phi$=0 need to be specified.  In this paper,
we define $\phi$=0 at periastron passage.  In most observational investigations 
$\phi_{obs}$=0 is defined at the orbital latitude  90$^\circ$ from the line of nodes.  This
definition is such that $\phi_{obs}$=0 at the midpoint of the X-ray eclipse (Bildsten et al. 1997, van Kerkwijk
et al. 1995, Barziv et al. 2001), and periastron passage occurs at $\phi_{obs}$=0.17.  Hence, $\phi_{obs}$=$\phi-$0.17.

\section{Observationally constrained parameters of GP Vel}

The spectral type and class of the optical component in GP Vel is given by Morgan et al. (1955) as
B0.5Ib. Howarth et al. (1997) assign the same spectral type but a more luminous  class,  B0.5Iae. 
Estimates of $m_1$'s mass lie in the range 21.7 M$_\odot$ (van Paradijs et al. 1977a, assuming $i$=78.8$^\circ$)
to 24.0 M$_\odot$ (Rawls et al. 2011).  Van Kerkwijk et al. (1995) derive  a value for the stellar radius, 
$R_1$=29.9--30.2 R$_\odot$, under the assumption that the supergiant star fills its effective Roche lobe at
periastron.  Rawls et al. (2011) derived $R_1$=31.82, under the same assumption. 
These values are consistent with the radii of B0.5Ia (26--38 $R_\odot$) stars listed in 
the catalogue of Passinetti-Fracassini et al. (2001), but not with those of B0.5Ib (18-26 $R_\odot$) stars.  
Hence,  GP Vel's radius is quite uncertain but most likely lies in the range 26--32 R$_\odot$. 
Its effective temperature, $T_{eff}$=25000$\pm$1000 K was determined by Sadakane 
et al. (1985)  from the equivalent width of UV Fe IV and Fe III lines, and a comparison with other 
B0-B1 Ib stars. Fraser et al. (2010) determined log($g$/cm s$^{-2}$)=2.90 and  $T_{eff}$=26500 K from a 
model atmosphere fit to the spectrum.   

The orbital inclination angle  depends on GP Vel's assumed radius.  Given the large range
of possible values for the radius, it is, hence, rather uncertain. It  also depends  on the observed 
duration of the X-ray eclipse, $\theta_c$, which, however, is time variable and energy dependent 
(Quaintrell et al. 2003).  With these considerations in mind,  most authors agree that $i\geq$78$^\circ$.  
A recent analysis  by Rawls et al. (2011) yields as most probable values $i$=85.9$^\circ$  and
$i$=78.8$^\circ$.\footnote{This is the value given in their Table 4, although in the text the value 
quoted is 77.8$^\circ$.} 

Table 2 summarizes the ranges in the parameters that are generally adopted for the GP Vel system.
The most reliable of its parameters are those that have been determined from the X-ray pulsar's delay 
times; specifically, $P_{orb}$,  $e$,  the projected semi-major axis of the neutron star orbit, $a_X$sin$i$,
and argument of periastron, $\omega_{per}$.  We adopt these as the basic known parameters
and keep them fixed throughout this paper.  

Given the considerable uncertainties that are associated with the remaining parameters, we opted to 
construct sets of self-consistent parameters for conducting the numerical simulations. 
The method for deriving the self-consistent parameter sets is described in sections 3.1--3.3. A sample of  
self-consistent parameters is listed in Table 3.  The TIDES code was run first for a selection of these  
parameters.  We then constructed a second block of parameters with small variations of one or more of 
the self-consistent parameters and the TIDES code was run for these as well.


\begin{table}     
\caption{Published stellar and orbital parameters.\label{tbl-1}}
\label{table1}
\centering
\begin{tabular}{lcll}
\hline\hline
Parameter          &         Value    &   & Reference                              \\
\hline
P (days)           & 8.964368         &   & Bildsten et al. 1997                   \\
e                  & 0.0898           &   & Bildsten et al. 1997                  \\
$a_X$sin$i$(ls)    & 113.89$\pm$0.13  &   & Bildsten et al. 1997                   \\
$\theta_e (^\circ)$& 30--36           &   & van Kerkwijk et al. 1995               \\
$\omega_{per}(^\circ)$ & 332.59$\pm$0.92& & Bildstein et al. 1997                  \\
\hline
$\vv$sin$i$ (km/s) & 116$\pm$6        &   & Zuiderwijk 1995                        \\
                   &  56$\pm$10       &   & Fraser et al. 2010                     \\
\hline
$K_{opt}$ (km/s)   &   17--29.7       &   & van Kerkwijk et al. 1995               \\
                   &  21.7$\pm$1.6    &   & Barziv et al. 2001;                    \\
                   &  22.6$\pm$1.5    &   & Quaintrell et al. 2003                 \\
\hline
log~$g$ (cm/s$^2$)  & 2.90$\pm$0.2     &   & Fraser et al. 2010                     \\
\hline
$T_{eff}$ ($^\circ$K)& 26500          &   & Fraser et al. 2010                     \\
                   & 25000$\pm$1000   &   &Sadakane et al. 1985                    \\
\hline
$V_\infty$ (km/s)  & 1105             &   & Howarth et al. 1997                    \\
\hline
$i (^\circ)$       &   78.8           &   & Rawls et al. 2011                      \\
                   & 70.1-90.         &   & Quaintrell et al. 2003                \\
\hline
$M_1$   ($M_\odot$)& 23.2-23.6        &   & van Kerkwijk et al. 1995               \\
                   & 24.00            &   & Rawls et al. 2011                      \\
                   & 23.8$^{+2.4}_{-1.0}$&& Barziv et al. 2001                     \\
                   & 23.1-27.9        &   & Quaintrell et al. 2003                 \\
\hline
$R_1$ ($R_\odot$)  & 31.82            &   & Rawls et al.         \\
                   & 29.9-30.2        &   &  van Kerkwijk et al. 1995  \\
                   & 30.4$^{+1.6}_{-2.1}$&& Barziv et al. 2001                     \\
                   & 26.8-32.1        &   & Quaintrell et al. 2003                 \\
\hline
\hline
\end{tabular}
\end{table}


\subsection{Stellar masses}

The mass ratio, $q=m_2/m_1$, is constrained by the known value of $a_X$~sin$i$ and Kepler's
Third Law $P^2 = \frac{4 \pi}{G(m_1 + m_2)} a^3$.
Using $q=m_2/m_1=a_1/a_2$, and  and  $a=a_1 + a_2=a_2(1+q)$, with $a_1$ and $a_2=a_X$ the semimajor 
axes  of the two stars' orbits, 

\begin{equation}
q=\frac{P}{2 \pi} \left(\frac{G m_1~sin^3i}{(a_X~sin i)^3}\right)^{1/2}-1  
\end{equation}

\noindent  Adopting $P$=8.964368, $a_x sin i$=113.89 light-seconds (Bildsten et al. 1997),
and writing $m_1$ in solar units,

\begin{equation}
q=0.22505 (sin~i)^{3/2} ~ (m_1/M_\odot)^{1/2} - 1
\end{equation}

Thus, for a fixed value of $i$,  each value of $m_1$ defines a unique value of $m_2$. The range in 
$m_1$=22.5--24.5 M$_\odot$ leads to a range $m_2$=1.427--2.04 M$_\odot$ for the cases analyzed for
$i$=78.8$^\circ$.



\subsection{Primary Radius, $R_1$}
The value of the B-supergiant's radius, $R_1$, is particularly relevant  since the tidal forces scale as $R_1^{-3}$.
There are three constraints on its value.  The first relies on the assumption that it is close to
filling its Roche Lobe, an assumption that requires knowledge of its rotation velocity and the
neutron star's mass, $m_{ns}$.  The second relies on: 1) the duration of the X-ray eclipse, 2) the
orbital separation, $a$, and 3) the orbital inclination, $i$.   
The third constraint on $R_1$ relies on a stellar atmosphere model fit to
its spectrum which yields $log~g$, from which $R_1$ can be derived given knowledge of $m_1$.
We have chosen the latter constraint to fix $R_1$ values for the calculations.  With the
observational constraint on $log~g$, the value of  $R_1$ follows from the choice of $m_1$  using 
$ R_1=\left( \frac{G m_1}{g} \right)^{1/2}$, where $g$ is the value of the surface gravity.
With $log~g=$2.90 cm s$^{-2}$ from Fraser et al. (2010),

\begin{equation}
R_1=5.8396~(m_1/M_\odot)^{1/2} R_\odot
\end{equation}

For the range in $m_1$ that was analyzed, the above expression leads to a range $R_1$=27.7--28.9 R$_\odot$.
It is important to note that $R_1$ is the equilibrium radius of $m_1$ in the absence of $m_2$ and
assuming no rotational deformation. These two effects lead to an aspherical shape and a 
unique value of the stellar radius cannot be defined. 

\subsection{RV curve semi-amplitude, $K_1$}

The orbital velocity  semi-amplitude is given by  Binnendijk (1960, p. 151): 

\begin{equation}
K_1=\frac{2 \pi~a_1~sin~i}{P(1-e^2)^{1/2}}  
\end{equation}

\noindent Writing $a_1 = a/(1+\frac{m_1}{m_2})$, and again using Kepler's Third Law,  

\begin{equation}
K_1= (2 \pi G)^{1/3} \frac{sin i}{(1-e^2)^{1/2}} m_2^{1/3} P^{-1/3} \left(1+\frac{m_1}{m_2}\right)^{-2/3}
\end{equation}

\noindent Determinations of $K_1$ since 1976  lie in the range $\sim$17--28 km/s (see Table 1 of 
Quaintrell et al. 2003 for a summary).  More stringent limits have been given  by Quaintrell et al. (22.6 $\pm$1.5 km/s)
and Barziv et al. (21.7 $\pm$1.6 km/s).  These limits are derived from fits of Keplerian RV curves to the 
observational data.

\subsection{Equatorial rotation velocity}

There are two widely different determinations of the projected equatorial rotation velocity, 
$\vv~sin~i$.  Zuiderwijk (1995) obtained $\vv~sin~i$=116 $\pm$6 km/s  by fitting 
synthetic profiles  to the observations.  This is consistent with the 114 km/s 
obtained by Howarth et al. (1997).  Recently, however, Fraser et al. (2010) applied the 
Fourier transform method (Gray 2008) to the spectrum and obtained $\vv~sin~i$=56 km/s.  
They attribute the additional broadening observed in the spectral lines to ``macroturbulence". 

It is important to note that, in general,  the stellar rotation velocity  plays  
a very important role in determining the behavior of  the stellar surface.  For example,  
faster rotation leads to a larger deformation of the star.  In addition, the ratio of stellar 
angular rotation velocity, $\omega$, to orbital angular velocity, $\Omega$, plays a critical 
role in the time dependence and amplitude of the tidal forcing.  Thus, the  synchronicity  parameter 
$\beta=\omega/\Omega$, which describes the degree of departure from an equilibrium configuration, is of
importance in the calculation. Note that  when $\beta$=1, the system is in synchronous rotation, which 
is one of the conditions for equilibrium. In an eccentric binary, $\Omega$ changes with orbital phase 
and thus, $\beta$ is a function of orbital phase.

In our model, the stellar rotation velocity is specified as an input parameter through $\beta_0=\omega/\Omega_0$,
where $\Omega_0$ is the orbital angular velocity at periastron. It can be conveniently expressed as,

\begin{equation}
\beta_0=0.02 \frac{\vv_{rot}/km s^{-1}}{R_1/R_\odot} (P_{orb}/days) \frac{(1-e)^{3/2}}{(1+e)^{1/2}}
\end{equation}

The two different values of $\vv~sin~i$ imply   $\beta_0\sim$ 0.3 or 0.6, (for $R_1\sim$28--29 R$_\odot$ and
$i >$78$^\circ$).
We shall refer to the case $\beta_0 \sim$0.6 as the ``standard" case and the $\beta_0 \sim$0.3 as the ``slow" 
rotation case.

\begin{table}
\begin{center}
\caption{Sets of self-consistent input parameters.\label{table-3}}
\centering
\begin{tabular}{lllllllll}
\hline\hline
 m1 & m2 & i  &  $R_1$ & $\vv_{rot}^s$ & $\vv_{rot}^f$ &$\beta_0^s$  & $\beta_0^f$  &K$_{Kep}$      \\
\hline
 23.6 & 1.468 &78.8 &28.36 & 57.09 &118.25  & 0.300&0.622 &17.37    \\
 24.0 & 1.708 &78.8 &28.60 & 57.09 &118.25  & 0.298&0.617 &19.87    \\
 24.2 & 1.830 &78.8 &28.72 & 57.09 &118.25  & 0.296&0.614 &21.12    \\
 22.5 & 1.427 &85.9 &27.69 & 56.14 &116.30  & 0.302&0.626 &17.71     \\
 22.7 & 1.546 &85.9 &27.82 & 56.14 &116.30  & 0.301&0.624 &19.02    \\
\hline
\end{tabular}
\end{center}
This table lists a sample of self-consistent input parameters derived
as described in Section 3.  Cols. 1 and 2 list the stellar masses,
in M$_\odot$; col. 3 the orbital inclination; col. 4 the stellar radius,
in R$_\odot$, cols. 5 and 6 the ``slow" and the ``fast"  rotation velocities,
in km/s; cols. 6 and 7 the corresponding asynchronicity parameters; and col. 8
the semi-amplitude of the Keplerian radial velocity curve, in km/s.
\end{table}

\begin{figure}
\centering
\includegraphics[width=7cm]{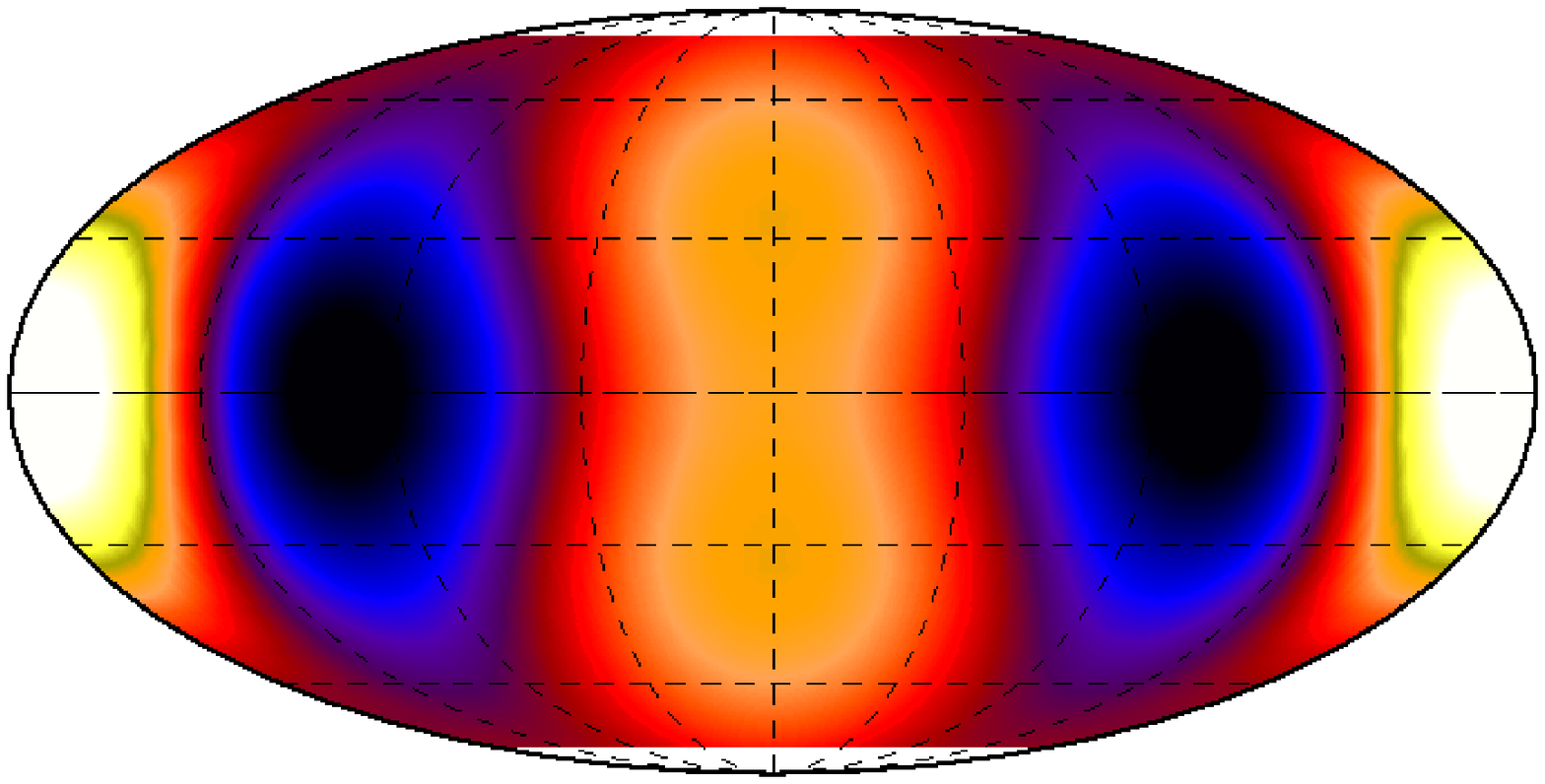}
\includegraphics[width=8cm]{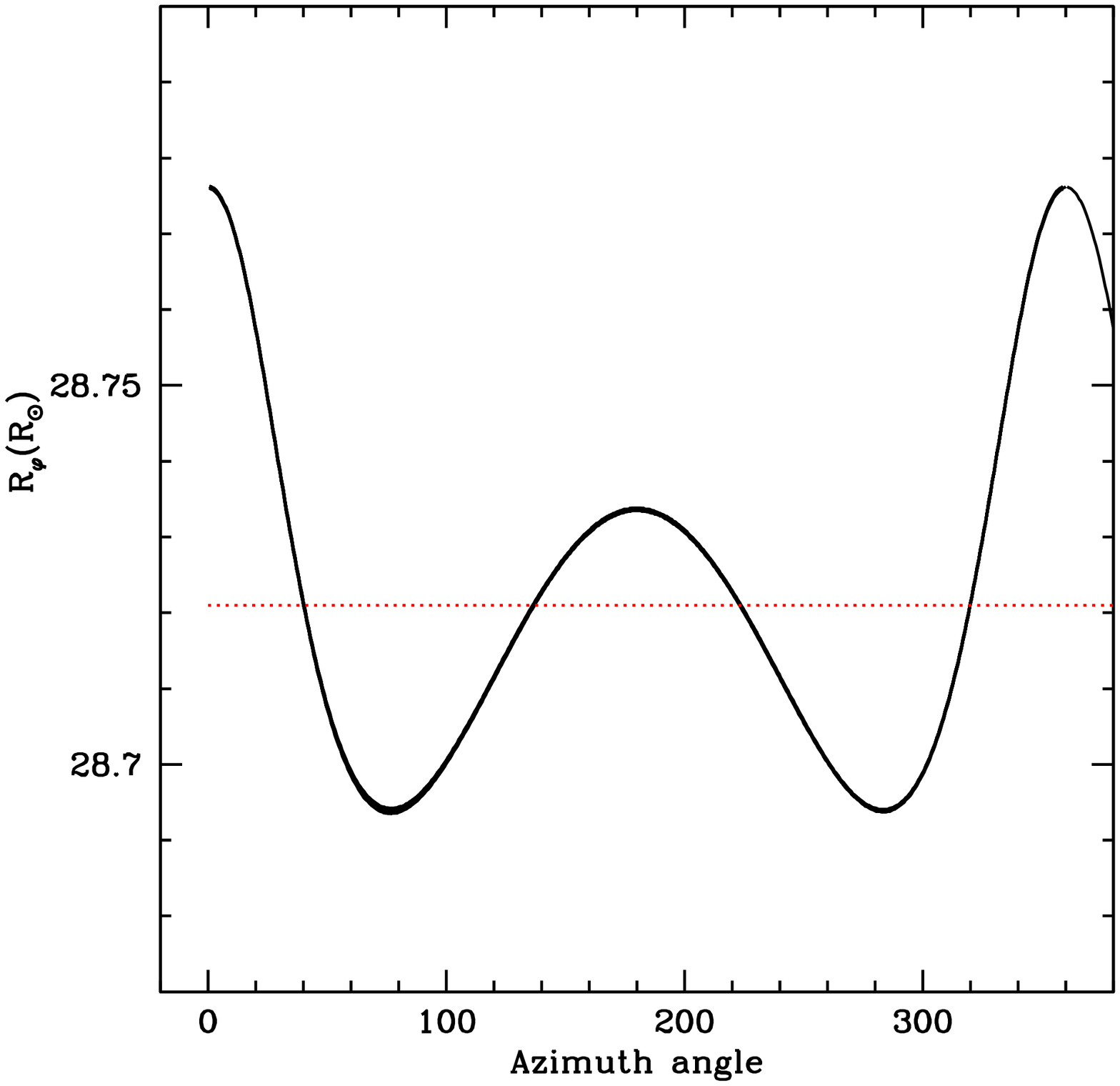}
\caption{Top: Shape of the stellar surface in  the equilibrium configuration (case 065); black/white color coding
corresponds to the minimum/maximum radius; the map is centered on 180$^\circ$ longitude.  The asymmetry in the
tidal bulges stems from the fact that the orbital separation is small compared to the radius of $m_1$. Bottom:
The radius as a function of longitude at the equator.  All the 40 orbital phases for which it was computed
are plotted, showing that no surface perturbations are present in this calculation.
}
\end{figure}


\onecolumn
\begin{table}[!h,!t,!b]
\begin{center}
\caption{Model input parameters and RV curve amplitudes. \label{tbl-2}}
\centering
\begin{tabular}{lllllllllllll}
\hline\hline
Case   &  $m_1$ & $m_2$ & $i$  & $\beta_0$& $dR/R_1$  & $R_1$ & $\nu$ &  K$_{fit}$  &K$_{Kep}$    \\ 
\hline
61a    & 23.6&1.468&78.8&0.300&0.06&28.363&0.12&              18.37  &17.37              \\   
61b    & 23.6&1.468&78.8&0.300&0.10&28.363&0.12&              20.14  &17.37              \\
61c    & 23.6&1.468&78.8&0.622&0.06&28.363&0.12&              17.70  &17.37              \\   
61d    & 23.6&1.468&78.8&0.622&0.10&28.363&0.22&              17.26  &17.37              \\   
64d    & 24.0&1.708&78.8&0.617&0.10&28.602&0.22&              20.01  &19.87              \\     
65a    & 24.2&1.830&78.8&0.296&0.06&28.721&0.22&              22.05  &21.12              \\   
65b    & 24.2&1.830&78.8&0.296&0.10&28.721&0.22&              24.37  &21.12              \\  
65c    & 24.2&1.830&78.8&0.614&0.06&28.721&0.22&              21.56  &21.12              \\   
65d    & 24.2&1.830&78.8&0.614&0.10&28.721&0.22&              21.96  &21.12              \\   
65e    & 24.2&1.830&78.8&0.614&0.08&28.721&0.22&              21.51  &21.12                \\    
65f    & 24.2&1.830&78.8&0.614&0.06&28.721&0.22&              21.34  &21.12              \\
71a    & 22.5&1.427&85.9&0.302&0.06&27.694& 0.12&             18.62   &17.71              \\   
71b    & 22.5&1.427&85.9&0.302&0.10&27.694& 0.12&             20.76   &17.71              \\   
71c    & 22.5&1.427&85.9&0.626&0.06&27.694& 0.12&             17.88   &17.71              \\   
71d    & 22.5&1.427&85.9&0.626&0.10&27.694& 0.12&             17.99   &17.71              \\
72a    & 22.7&1.546&85.9&0.301&0.06&27.817& 0.22&             20.17   &19.03              \\    
72b    & 22.7&1.546&85.9&0.301&0.10&27.817& 0.22&             22.11   &19.03              \\   
72c    & 22.7&1.546&85.9&0.624&0.06&27.817& 0.22&             19.34   &19.03              \\   
72d    & 22.7&1.546&85.9&0.624&0.10&27.817& 0.22&             19.38   &19.03              \\   
72e    & 22.7&1.546&85.9&0.296&0.10&28.3  & 0.22&             22.45   &19.03              \\
72f    & 22.7&1.546&85.9&0.613&0.10&28.3  & 0.22&             19.49   &19.03              \\ 
30     & 23.5&1.44&78.8&0.301&0.06 &28.3&0.22 &               18.29 &17.2                  \\  
31     & 23.5&1.44&78.8&0.301&0.10 &28.3&0.22 &               20.33 &22.1                  \\
32     & 23.5&1.44&78.8&0.622&0.06 &28.3&0.22 &               17.32 &17.2                  \\
33     & 23.5&1.44&78.8&0.622&0.10 &28.3&0.22 &               17.06 &17.2                  \\
34     & 23.5&1.44&78.8&0.300&0.06 &28.8&0.22 &               18.05 &18.05                \\ 
34b    & 23.5&1.44&78.8&0.300&0.10 &28.8&0.22 &               19.90 &18.05               \\
34c    & 23.5&1.44&78.8&0.620&0.06 &28.8&0.22 &               17.55 &18.05               \\
34d    & 23.5&1.44&78.8&0.620&0.10 &28.8&0.22 &               17.38 &18.05               \\
35     & 24.0&1.74&78.8&0.298&0.06 &28.6&0.22 &               21.10 &20.4                  \\
36     & 24.0&1.74&78.8&0.298&0.10 &28.6&0.22 &               22.98 &20.4                  \\
37     & 24.0&1.74&78.8&0.615&0.06 &28.6&0.22 &               20.64 &20.4                  \\
38     & 24.0&1.74&78.8&0.615&0.10 &28.6&0.22 &               20.55 &20.4                  \\
41     & 24.5&2.04&78.8&0.609&0.10 &28.9&0.22 &               23.50 &23.4                  \\
50     & 22.5&1.52&90.0&0.307&0.06 &27.7&0.12 &               19.99 &19.5               \\   
51     & 22.5&1.52&90.0& 0.307&0.10 &27.7&0.12&               22.25 &19.5               \\   
52     & 22.5&1.52&90.0& 0.624&0.06 &27.7&0.12&               19.14 &19.5               \\    
53     & 22.5&1.52&90.0& 0.624&0.10 &27.7&0.12&               19.35 &19.5               \\    
55     & 23.5&1.44&85.9& 0.301&0.06 &28.3&0.12&               18.38 &18.05              \\   
56     & 23.5&1.44&85.9& 0.301&0.10 &28.3&0.12&               20.60 &18.05              \\
57     & 23.5&1.44&85.9& 0.622&0.06 &28.3&0.12&               17.65 &18.05              \\  
58     & 23.5&1.44&85.9& 0.622&0.10 &28.3&0.12&               17.29 &18.05              \\   
\hline
\hline
\end{tabular}
\end{center}
This table lists the input parameters for the models that were computed. Cases 61a--72d 
were computed with ``self-consistent" sets of input parameters, as described in Section 3.
$m_1$, $m_2$ are given in
in Solar masses; $R_1$ in Solar Radii; $\nu$ in R$_\odot^2$/day; and  K$_{fit}$ and K$_{Kep}$ in
km/s.
\end{table}
\twocolumn
\section{Results}

Table 4  lists the input parameters of the model runs performed for this paper. The first block
of models (cases 61a--72d) are run with self-consistent sets of parameters, derived as is described in the 
previous section.  The parameters of the second block (list starting with case 72e)  are not fully self-consistent.
Columns 2--5, and column 7 list, respectively, $m_1$, $m_2$, $i$, $\beta_0$ and $R_1$; column 6 contains 
the depth of the surface layer, $dR/R_1$, used in the model.  Columnn 8 lists the value of the kinematical 
viscosity, $\nu$ (in units of $R_\odot^2$/day).  

\subsection{The equilibrium case}

It is illustrative to first analyze an equilibrium case having parameters similar to
those of the  GP Vel system, but in which $e$=0 and $\beta$=1.
  We chose the case with $m_1$=24.2 M$_\odot$, $m_2$=1.83  M$_\odot$,
$R_1$=28.721  R$_\odot$ and $\vv_{rot}$=118 km/s, the latter corresponding to the
``standard" rotation speed and an orbital inclination $i$=78.8$^\circ$.  In order
for this system to be in equilibrium (i.e., $\beta$=1), we had to set the orbital
period to 12.2 days.  We prefered modifying  $P$ instead of $\vv_{rot}$ because 
in order to make $\beta_0$=1 using GP Vel's orbital period, we would need to set 
$\vv_{rot}\sim$160 km/s, significantly larger than its actual rotation speed.  This
leads to a stronger deformation of the star.  The main difference of using a larger orbital
period instead of the actual period is that the tidal force is weaker, so the tidal deformation 
is smaller.  

The start of  the TIDES calculation is characterized by a transitory phase during which
the star adjusts from its initially spherical shape to the new equilibrium shape.  
During this transitory phase, the star undergoes initially large amplitude ($\sim$0.03 R$_\odot$ 
in the present case) oscillations that rapidly damp out, reaching an amplitude $<$0.002 
R$_\odot$ by 20 cycles after the start of the calculation.  At this time, the star has 
attained its equilibrium shape,  which consists of two tidal bulges.  Figure 1 shows a color-coded 
Mollweide representation of the stellar radius at each surface element.    The left 
edge of the map corresponds to azimuth angle 0$^\circ$ (i.e., the sub-binary longitude,
defined as the longitude that intersects   the line connecting the centers of the two stars) 
and the  right edge is at 360$^\circ$. The plot below the map illustrates the
magnitude of the radius at  the equatorial latitude as a function of azimuth.  Note that the  
size of the two tidal bulges is not the same  due to the fact that the orbital 
separation is relatively small compared to the stellar radius.  The height of the primary bulge 
above the equilibrium radius in this calculation is $\delta R$=0.06 R$_\odot$. This corresponds to 
$\delta R/R_1$=0.002, which is significantly smaller than the depth of the surface layer used in the TIDES 
calculation, $dR/R_1$=0.06.  Also, 
note that the primary bulge points directly towards the companion, another indication that the
system is in the equilibrium configuration. In non-equilibrium configurations, the bulge
either ``lags" behind or is advanced with respect to the line connecting the centers of the stars,
a phenomenon that leads to the tidal torques in the system.

Once the equilibrium configuration is attained, there are no significant surface motions.
Figure 1 (bottom) is actually a plot of the equatorial latitude of the star at the 40 orbital phases
for which it was computed.   The width of the curve is an indication of the stability over time 
of the configuration.  As a consequence,  no significant variability in the photospheric 
absorption-line profiles is  observed in our calculations.\footnote{Note that our model does not 
take into account a variable temperature structure across the observable
stellar disk.  In the case of a more complete calculation in which a model stellar atmosphere is
used, a phase-dependent variation in the line-profiles is expected due to gravity darkening
effects, see Zuiderwijk et al. 1977; and  Palate \& Rauw, 2011.}

\begin{figure}
\centering
\includegraphics[width=8cm]{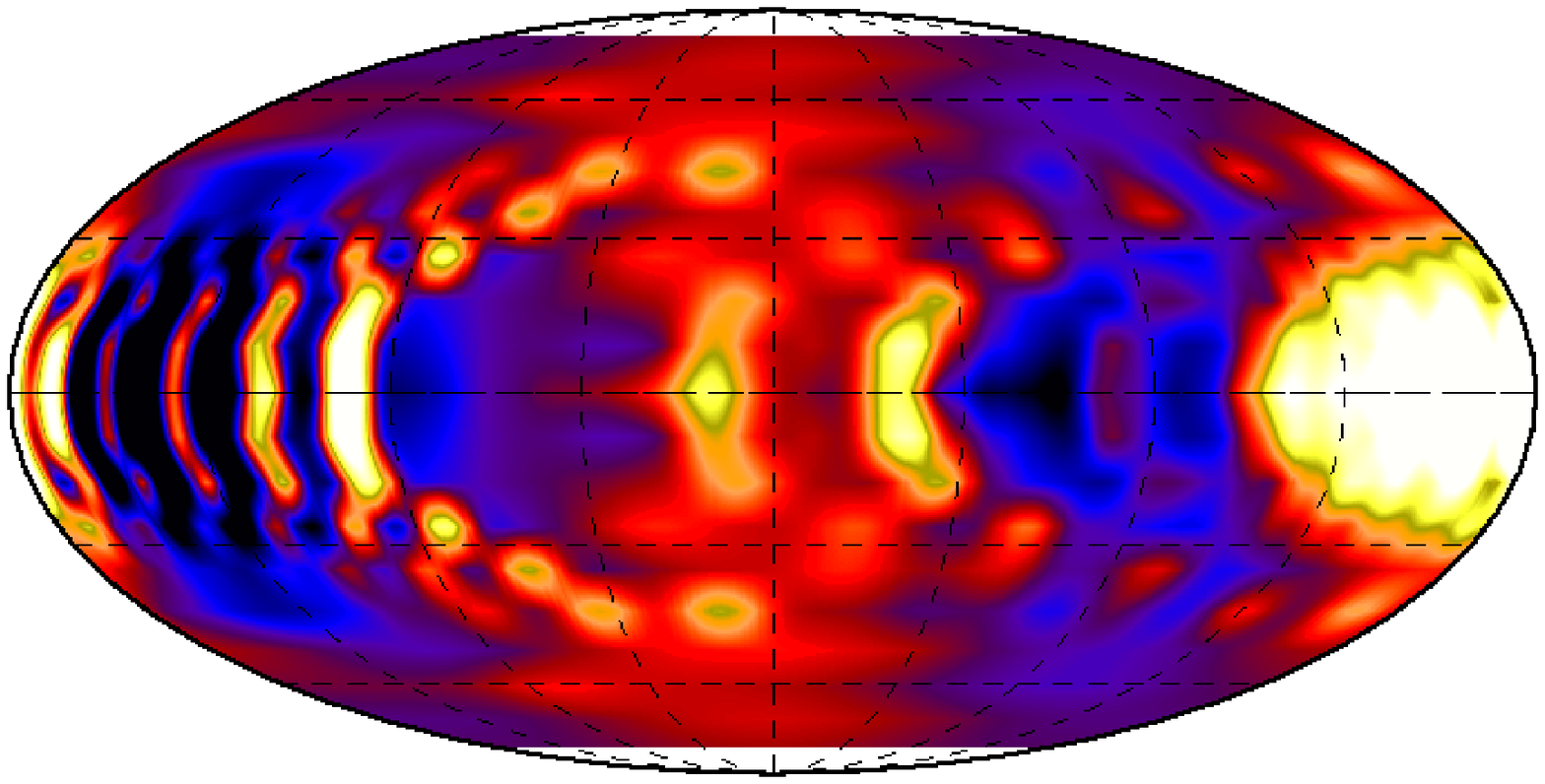}
\includegraphics[width=8cm]{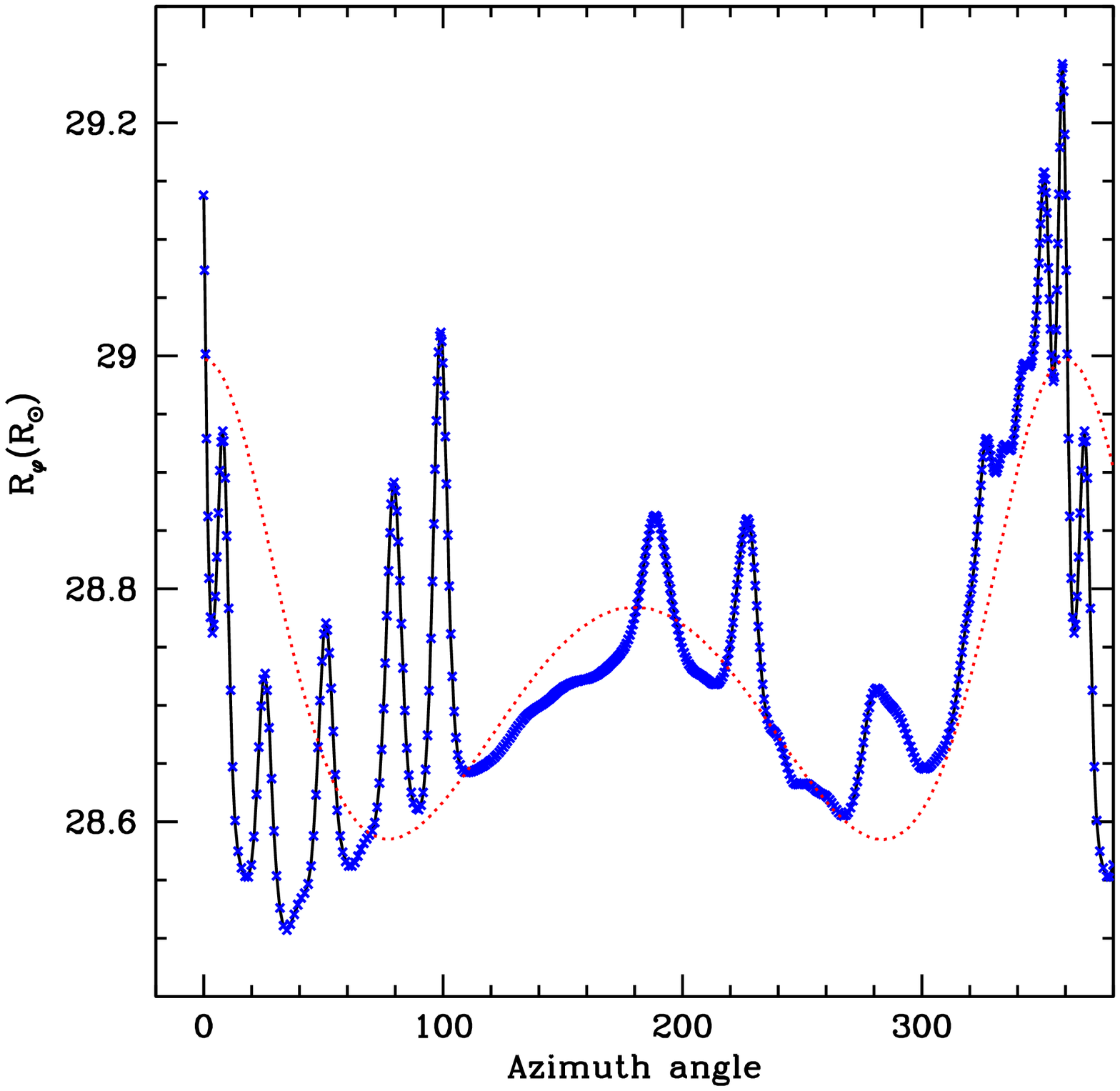}
\caption{Top:  Shape of the stellar surface  for ``standard" cases (model 65c) at periastron. 
Color coding and orientation are the same as in Figure 1. Bottom: The crosses represent the radius at
each azimuth angle along  the equator. The dotted line is the equilibrium
shape shown in Figure 1, rescaled by a factor of 3.  The strong departures from the
equilibrim configuration when $\beta_0$=0.6  are evident. The detailed pattern
changes as a function of orbital phase, causing the line-profile variations.
}
\end{figure}

\subsection{The ``standard" case}

The situation is very different when the system departs from the equilibrium configuration.
The fact that $\beta_0\neq$1 implies that the star rotates asynchronously, and this introduces
perturbations on the stellar surface. We illustrate this case with a model for 
$m_1$=24.2 M$_\odot$, $m_2$=1.83 M$_\odot$, $R_1$=28.721 R$_\odot$ (case 65c). Contrary to the 
equilibrium case, the  stellar surface does not attain a simple shape with two tidal bulges.  The 
map in Figure 2 shows a color-coded representation of the stellar radius at each surface element, 
with white/black indicating maximum/minimum extent. The map corresponds to the time of periastron 
passage.  The Mollweide representation and orientation are the same as in Figure 1. The primary tidal 
bulge  can be seen to lie between 320$^\circ$ and 360$^\circ$, ``lagging" behind the line connecting the
centers of the two stars (which lies at azumuth 0$^\circ$).  This is as expected for a 
sub-synchronously rotating star. Furthermore, the bulge shape is not smooth,  but consists of 
smaller-scale structures.  These can be most easily visualized in the plot shown below the map,
which shows the radius at the equator  plotted as a function of azimuth.  A similar configuration 
obtains for all other orbital phases as well.  The comparison between Figure 2 and the equilibrium 
configuration shown in Figure 1  shows that an ``equilibrium tide" representation for GP Vel is a
poor approximation. 

The non-equilibrium configuration leads to large-scale flows on the stellar surface (see Tassoul 
1987 and Eggleton et al. 1998 for a good description of this phenomenon). These are referred to 
as ``tidal flows".  The tidal flows are  motions of localized regions of the stellar surface relative 
to the underlying,  (assumed) rigidly-rotating stellar interior.   The changing flow patterns lead 
to variable line-profiles.  A sample of the type of variability predicted by the TIDES code is illustrated 
in Figure 3, where the perturbed line profiles are compared to  their corresponding non-perturbed 
line profiles.  Note the appearance of ``bumps" and ``wiggles" in the profiles, as well as the
occassional blue or red extended wing.  In calculations performed with significantly
smaller ``turbulent" speeds (for example, 10 km/s instead of the 30 km/s used here), the
``bumps" are narrower, and the profiles give the appearance of having narrow discrete 
absorption features that generally travel  from ``blue" to ``red" along the line profile
(see Harrington et al. 2009 for an example of this type of behavior).  In the case shown
in Figure 3 the discrete absorptions are significantly smoothed out due to the large
``turbulent" speed.

\begin{figure}
\centering
\includegraphics[width=8cm]{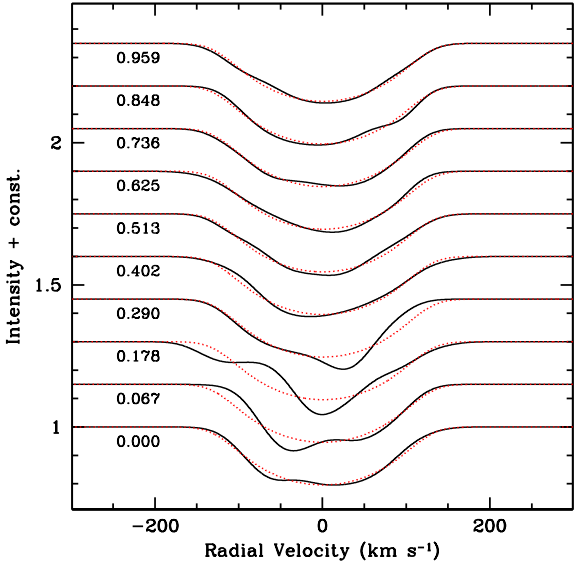}
\caption{Predicted line profiles  at different orbital phases for case 65c. The
dotted profiles correspond to the non-perturbed, rigidly-rotating surface.
The asymmetries in the perturbed line-profiles are responsible for the
departures from a Keplerian RV curve.
}
\end{figure}

\begin{figure}
\centering
\includegraphics[width=8cm]{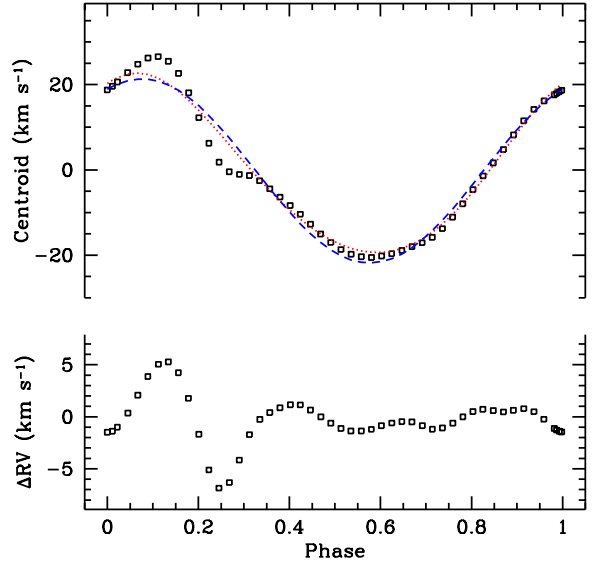}
\includegraphics[width=8cm]{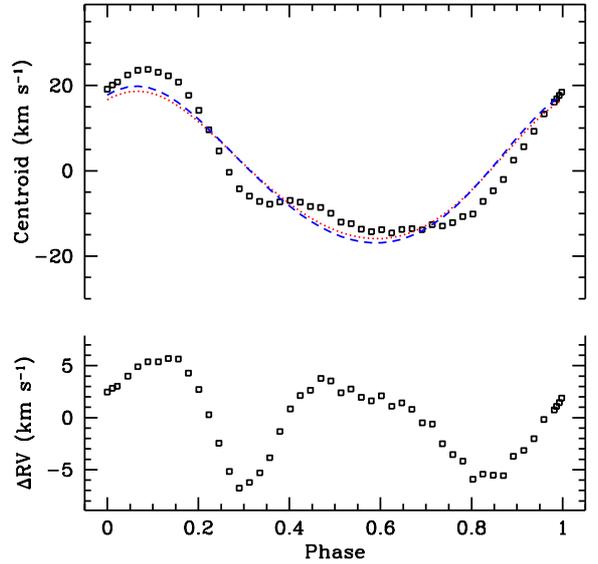}
\caption{Radial velocity curves from the  tidally perturbed line profiles (squares)
compared to the actual projected orbital motion  (dotted curve) and best-fit
curve (dash curve). $\Delta$RV is the difference between the perturbed and
non-perturbed RV values.
Top:  $m_{ns}$=1.830 M$_\odot$, $\beta_0$=0.614 (the "standard" case;
case65c); Bottom: $m_{ns}$=1.468 M$_\odot$, $\beta_0$=0.300 (case 61a).
}
\end{figure}

The RV curve derived from these line profiles, using the flux-weighted mean method, is plotted 
in Figure 4 (top).\footnote{A similar curve is derived if we plot the RV's obtained by fitting a
Gaussian to the line profiles}.   Also plotted in this figure is the  RV curve corresponding to the 
actual orbital motion (dots), and the best-fit Keplerian curve derived from the analysis that is 
described in Section 2.3 (dashes).  The deviations of the perturbed RV curve with respect to that
of the actual orbital motion are shown below the RV curve.   They are as 
large as $\pm$5 km/s in the phase intervals  $\phi$=0.1--0.3 and $\phi$=0.9--1.0.    
  Despite the large deformations on the perturbed RV curve, the best-fit Keplerian is very similar 
to the curve which describes the actual orbital motion.   Thus, although the peak-to-peak amplitude 
of the perturbed RV curve  is significantly larger than the actual orbital motion,
the ``excursions" of the RV curve are such that they nearly cancel out in the fitting algorithm.   
This indicates that Keplerian curve fits to the tidally-perturbed RV curves yield reasonable 
results  as long as the entire set of data is fit; i.e., {\it without excluding portions of the 
RV curve}.

The general characteristics described above appear in all of the  model runs listed in 
Table 4  for which $\beta_0\sim$0.6.  The semi-amplitudes of the corresponding RV curves derived
from the fits ($K_{lsq}$) are plotted as filled symbols in Figure 5 with $m_2$=$m_{ns}$ on the 
abscissa. Also plotted in this figure are the values of $K_1$ for the Keplerian orbit (crosses).
The observational limits are drawn with horizontal dotted lines.  Only models with 
$m_{ns}$=$m_2>$1.7 M$_\odot$ have values of $K_{lsq}$ that lie within the observational limits.  
Hence, we conclude that for the $\beta_0\sim$0.6 group of models, the systematic shifts in  RV 
caused by tidal flows produce only small departures from the actual orbital RV curve,  and 
the derived neutron star mass is not severely affected by the tidal flows.

\begin{figure}
\centering
\includegraphics[width=8cm]{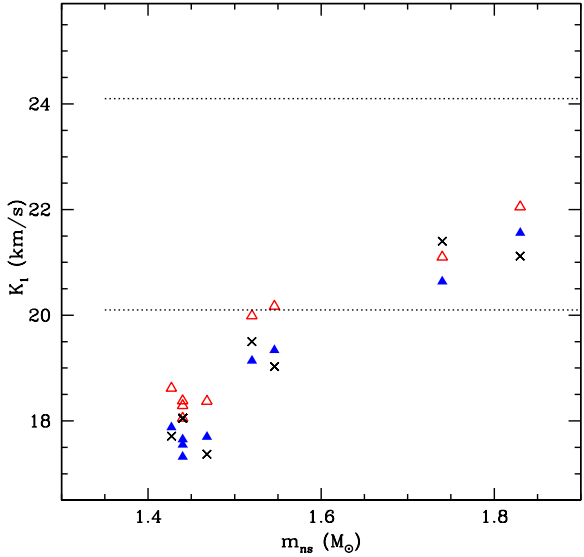}
\caption{Predicted semi-amplitude of the GP Vel RV curves 
for different assumed neutron star masses and model input parameters:
open symbols: $\beta_0\sim$0.3; filled symbols: $\beta_0\sim$0.6.  All models
were run with $dR/R_1$=0.06.
The dotted lines indicate the observed range of semi-amplitudes from Barziv et al. (2001)
and Quaintrell et al. (2003). The  Keplerian semi-amplitudes (crosses) are generally smaller
than those predicted by  the model calculations.  Note that  for the slow-rotation
($\beta_0\sim$0.3) models,  masses as low as $m_{ns}$=1.55 M$_\odot$  yield $K_1$
values within the observed range.
}
\end{figure}

\subsection{The ``slow" rotation case} 

The  slow rotation models also show  significant line-profile variability.
The deformation of the RV curves leads to systematically larger values of $K_{lsq}$.
To illustrate this case we  use a model with $m_1$=23.6 M$_\odot$, 
$m_2$=1.468 M$_\odot$, and $R_1$=28.363 R$_\odot$ (case 61a).  Figure 4 (bottom) shows that the 
perturbed RV curve departs significantly more from its  Keplerian motion than the 
$\beta_0 \sim$0.6 case.  There are  $\sim$5 km/s  excesses  around $\phi$=0.1 and 0.65,  phases 
which lie close to the extrema in the actual orbital  RV curve. This causes the best-fit RV 
curve (dashes) to have a significantly  larger amplitude than the curve describing the orbital motion (dots).

The summary of Keplerian fit  semi-amplitudes, $K_{lsq}$, for the $\beta_0 \sim$0.3 cases is shown
in  Figure 5  with open symbols.  For models in which the layer depth is $dR/R_1$=0.06, we see that
masses as low as $m_{ns}$=1.55 yield values of $K_{lsq}$ (marginally) consistent with the observations.

Thus, for the $\beta\sim$0.3 group of models, the systematic shifts in  RV caused by tidal flows produce 
significant deviations from the actual orbital RV curve.   


The basic conclusion of this section is that different rotation speeds lead to tidal flows with 
different characteristics and amplitudes and, thus, it is crucial to have a well-established value
for this parameter in order to properly estimate the  magnitude of the RV curve perturbations.


\subsection{Dependence on layer depth}

Calculations performed with different layer depths in the ``standard" rotation case yield similar
results.  However, significant differences are present in the ``slow" rotation cases where the 
$dR/R_1$=0.10 layer depths lead to larger RV curve amplitudes than those for $dR/R_1$=0.06.
Figure 6 illustrates the radial velocities obtained from calculations with these two layer depths.
Also plotted are the corresponding best-fit Keplerian curves
and the curve that describes the actual orbital motion (continuous curve).              

The largest amplitudes occur for $\beta_0\sim$0.3 and $dR/R_1$=0.10,  
as illustrated in Figure 7 where we plot $K_{lsq}$ as a function of $m_{ns}$.
These model calculations indicate that the $\beta_0\sim$0.3  systems with $m_{ns}$ as low 
as 1.47 M$_\odot$ are consistent with the observations.

\begin{figure}
\centering
\includegraphics[width=8cm]{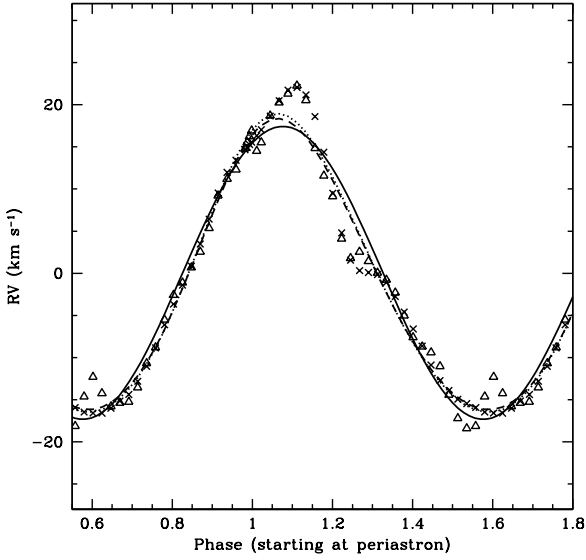}
\includegraphics[width=8cm]{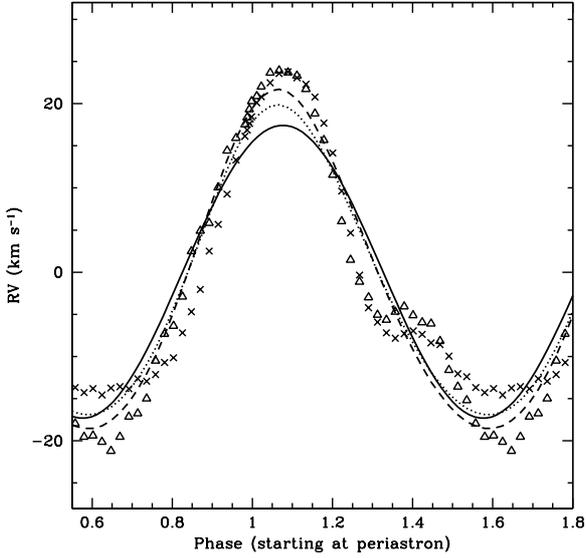}
\caption{Radial velocities obtained from calculations with different layer depths
for the ``standard" rotation velocity (top; cases 61c,d) and the ``slow" velocity
(bottom; cases 61a,b). Also plotted are the corresponding best-fit Keplerian curves
and the curve that describes the actual orbital motion (continuous curve).
Crosses and the dotted curve correspond to $dR/R_1$=0.06; and  triangles and the
dash curve correspond to $dR/R_1$=0.10.  The perturbations due to tidal flows are
more significant for the ``slow" rotation velocity cases.
}
\end{figure}

\begin{figure}
\centering
\includegraphics[width=8cm]{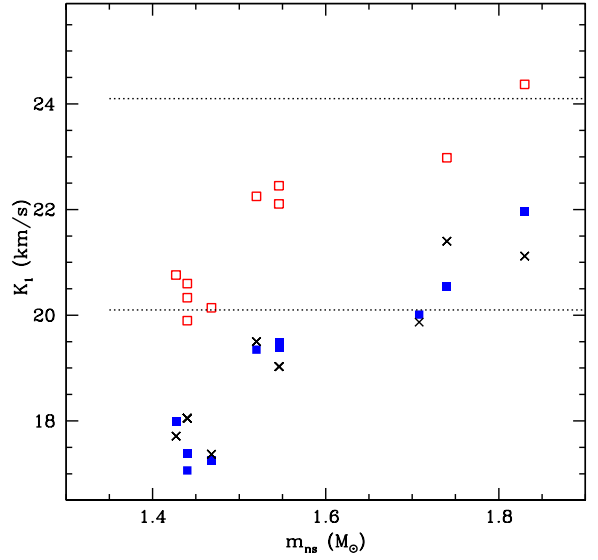}
\caption{Same as in Figure 5, but from calculations performed with a layer depth
$dR/R_1$=0.10.
}
\end{figure}

\subsection{The ``blue dip" and a structured wind}

A prominent feature in our model RV curves is a ``blue dip" that appears at $\phi\sim$0.3, shortly after
the maximum in the  curve. It is a persistent feature in all our computations, and it is caused by 
the asymmetrical shape of the line profiles. Figure 8 displays the line profiles over the phase interval 
0.12--0.36 for model 65c, illustrating the source of this feature.  Between phases 0.14--0.22, the line 
develops an excess blue absorption wing while at the same time the red wing becomes less extended.  The combination 
of these two effects moves the centroid of the line towards shorter wavelengths.  A natural question is what causes 
this asymmetry in the line profiles? 

A map of the azimuthal velocity perturbations over the visible surface of the B-supergiant at orbital phase 
$\phi$=0.2 is presented in Figure 9.  The light color indicates motion of the surface elements
that is faster than the underlying rigid-body rotation rate.  Hence,
the large white area on the left side of the map indicates that the surface elements that lie near
the limb of the star are approaching the observer faster than the stellar rotation velocity.  This leads to
the more extended blue wing in the line profiles.  Also evident in Figure 9 is the dark area in the
map, which corresponds to surface elements whose azimuthal velocity is slower than that of the stellar
rotation.  When this region reaches the right limb of the visible disk, it leads to
a less extended red wing in the absorption line profile.  

An extensive analysis of GP Vel's  observational RV curve has been made by Barziv et al. (2001) and 
Quaintrell et al. (2003), using independent data sets.  Barziv et al. (2001)  report
the presence of a ``blue excursion" in the H$\delta$ data that produces a local minimum in the RV curve
at $\phi\sim$0.37.  They interpret the ``blue excursion" in terms of a photoionization wake. 
Figure 10 is a plot of the H$\delta$ 4102 \AA\ line  RV values (crosses) from Barziv et al. (2001)
showing that the ``blue excursion" initiates at $\phi\sim$0.25.  This coincides  with the ``blue dip" in 
our models, but its duration is far longer than that of the ``blue dip".~    An alternative explanation
for the ``blue excursion" is the presence of  enhanced mass outflow after periastron passage 
and it is tempting to speculate that it could be triggered by the strong tidal effects that appear after 
periastron passage.

It is also interesting to note that Kreykenbohm et al. (2008) detected  flaring activity and temporary 
quasi-periodic oscillations in  INTEGRAL X-ray observations. In particular, the two largest flares observed 
in these data occur around orbital phase 0.4 (with respect to periastron) in two different orbital cycles.  
This is significant because  this is just  $\sim$0.03 in phase later than the minimum in the ``blue excursion" 
in Barziv et al.'s  data. Assuming a mean wind velocity of 1105 km/s (Howarth et al. 1997), a gas stream 
that originates near  the B-supergiant surface would travel  $\sim$3$\times$10$^{12}$ cm during this time interval,
which  is on the order of magnitude of the distance  to the neutron star.  Hence, the picture which emerges is
one in which tidal flows may cause instabilities that produce time-dependent ouflows leading to a structured
wind.  When high-density wind regions reach the neutron star, the larger accretion rates lead to an increase in the 
X-ray emission, as described by Kreykenbohm et al. (2008).
These authors also found temporary quasi-periodic oscillations with a timescale of $\sim$2 hr which they 
attribute to alternating high and low density regions within GP Vel's stellar wind.  A similarly  structured
wind has  been suggested to exist in the Be/X-ray binary system 2S0114+650 and to possibly be produced by 
tidally induced pulsations (Koenigsberger et al. 2006).


\begin{figure}
\centering
\includegraphics[width=8cm]{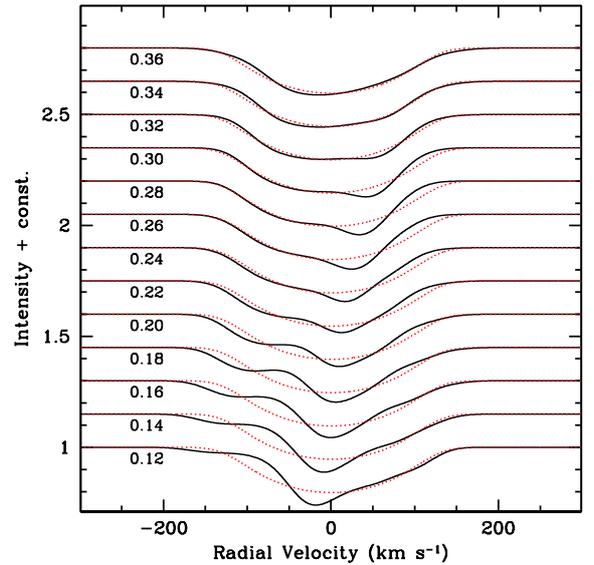}
\caption{Montage of line profiles at orbital phases $\phi$=0.12--0.36, showing
that the ``dip" in the RV curve is largely a consequence of the more
extended blue wing and less extended red wing.  Dotted curves are the unperturbed
line profiles.  Phases with respect to periastron passage are listed.
}
\end{figure}

\begin{figure}
\centering
\includegraphics[width=7cm]{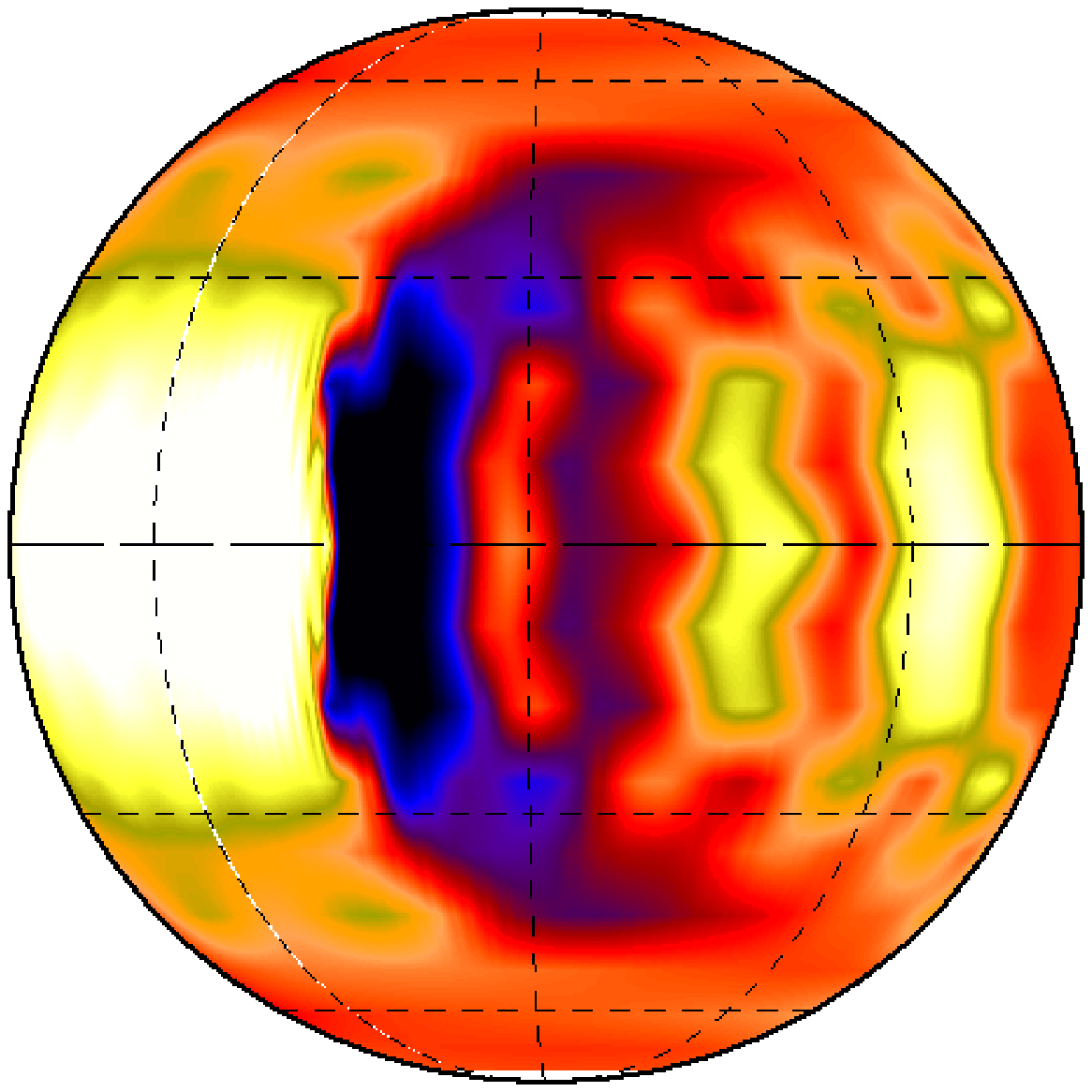}
\includegraphics[width=7cm]{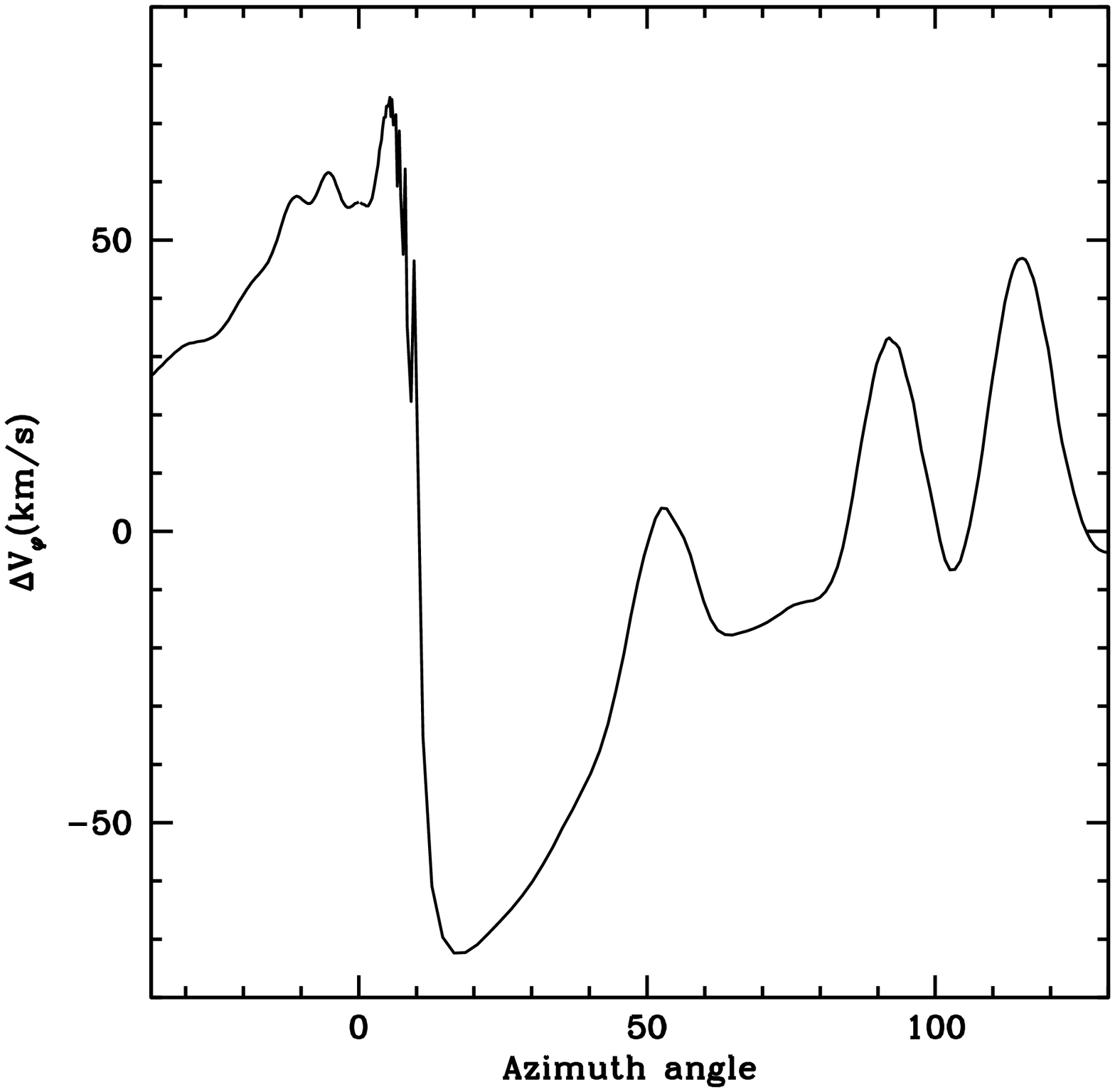}
\caption{Map depicting the azimuthal (``horizontal") velocity field at orbital
phase $\phi$=0.2 for case65c. Velocities that are larger than the rigid-body rotation rate
are coded in light color, while those that are slower are coded in dark.  The
``blue dip" in the theoretical RV curve is caused by the extended blue wing in the
photospheric absorption lines, and this is produced by the large white area near
the approaching (left) limb  of the star. The bottom panel is a plot of the azimuthal
velocity perturbations along the equatorial latitude.
}
\end{figure}

\begin{figure}
\centering
\includegraphics[width=8cm]{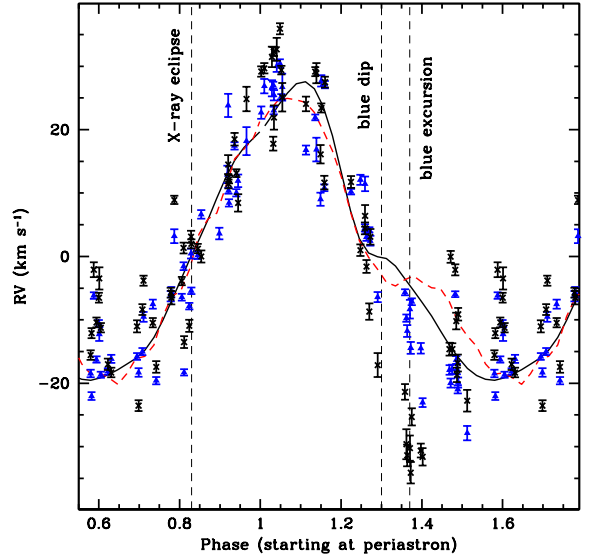}
\caption{Theoretical RV curves for a neutron-star mass  m$_{ns}=$1.83 M$_\odot$
for $\beta_0$=0.614 (continuous curve; case 65c) and m$_{ns}=$1.468 M$_\odot$
for $\beta_0$=0.300 (dash curve; case 61b). Data points are the  RV measurements
from Barziv et al. (2001): the average of several lines (triangles) and the values
for $H\delta$ (crosses).  The vertical dash lines indicate the phase of  X-ray eclipse,
the approximate location of the blue ``dip" in the RV curve and the ``blue excursion".  The latter may
be due to outflows from the B-supergiant.  We suggest these outflows may be associated
with activity induced by the tidal flows.
}
\end{figure}

\section{Discussion and Conclusions}

In this paper, we  explored the manner in which  surface motions that are induced by the tidal interactions affect
the shape  of the radial velocity curve. The objective was to  analyze the manner  in which  these motions  affect 
the neutron star mass determination. We performed a calculation from first principles which involves solving
the equations of motions of a Lagrangian grid of surface elements  to determine the surface velocity field on the
B-supergiant.  The velocities on the visible surface of the star were then projected along the line-of-sight to the
observer and  the local line-profile at  each  surface element was Doppler-shifted correspondingly.  The Doppler-shifted
local line-profiles were combined to obtain the absorption-line profile that would be detected by an observer at each
orbital phase. For each set of model binary parameters, the centroids of the resulting line-profiles were  measured
and a theoretical RV curve constructed.  

In all our models, the peak-to-peak amplitude of the model RV curve is larger than the corresponding Keplerian 
RV curve.  However,  the  fit  using Keplerian curves yields semi-amplitudes that may have only 
small departures from the true orbital motion, depending on the assumed stellar rotation rate. In the ``standard"
rotation case, the deformations of the RV curve are such that they nearly cancel out in the Keplerian fit.
Thus, the derived semi-amplitude of the RV curve in this case is only slightly larger than that of the actual orbital
motion. If GP Vel rotates at the ``standard" rotation rate, then the neutron star mass may indeed be $\geq$1.7 M$_\odot$.
However,  in the ``slow" rotation velocity cases,  the deformations of the RV curve are significantly more pronounced 
and  the Keplerian fits to the RV curves yield significanlty larger amplitudes than that of the orbital motion.  In
this case, $m_{ns}\sim$1.5 M$_\odot$ is feasible.


Thus, the assumed  rotation speed clearly plays a very important role in determining the effect of the tidal
flows on the amplitude of the  RV curve.  The ``standard" and ``slow" speeds we have adopted here are 
the two values cited in the literature.  However, because of the prominent line-profile variability, it is 
not clear whether either of these two values accurately represents the rotation velocity of the B-supergiant.   
This is an issue that requires further study.


The analysis of the RV curve is further complicated by the observed variations that occur from one orbit 
to the next. Trial calculations do show small differences in the predicted shape of the star for different 
orbital cycles, and the maximum extent of the primary tidal bulge varies from one periastron passage to 
the next.  However, these do not lead to significant differences in the predicted line-profiles at the 
same orbital phase  for different cycles, and thus the theoretical RV curves are virtually the same from 
one orbital cycle to the next.  

A final question to address is the following:  If the distortions in the radial velocity curve introduced by
tidal flows on the surface of GP Vel lead to an artificially large neutron star mass, shouldn't this 
same effect be present in other  binary systems ?  The answer is that, indeed, the effect should be 
present in binary systems in which the stellar rotation departs  from synchronous rotation.
In this context, it is interesting to consider the set of neutron star binary systems with optical 
counterparts which is listed in Table 5.  These are  the 7 known neutron stars with optical counterparts 
for which stellar parameters have been well determined.  Five of these have OB-type spectral types, while the
remainder have A--K spectral types.  The orbital periods range from $\sim$1 day to $\sim$10 days.
Figure 11 is a plot of the derived neutron star mass {\it vs.} ($1.-\beta_0$),  which tells us whether 
there is a dependence of $m_{ns}$ on the degree of departure of the optical component from synchronous rotation.  
A possible trend for larger $m_{ns}$ with larger departures from synchronicity is apparent in Figure 11, although
more data points are clearly required before we can conclude that a correlation exists.
However, this result is consistent with the idea that asynchronous rotation leads to
significant line-profile variability which, in turn, distorts the RV curve, potentially leading to an
error in the stellar masses.  Furthermore, this hypothesis applies to all types of binary systems.

\begin{figure}
\centering
\includegraphics[width=8cm]{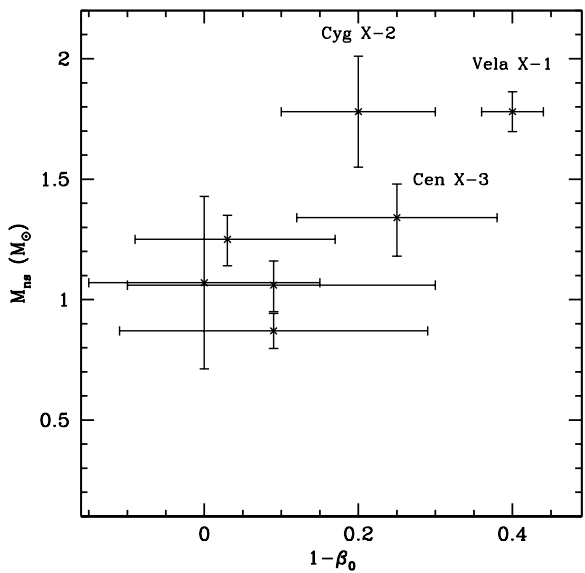}
\caption{The neutron star mass plotted {\it vs.} the asynchronicity factor
1-$\beta_0$ for the 7 binary systems listed in Table 5.  Synchronously-rotating
optical components lie at 1-$\beta_0$=0.  Error bars are the uncertainties quoted
by the authors listed in Table 5.  A trend for more massive neutron stars with
larger departures from synchronicity is observed.  Whether this indicates a
systematic error in the mass determination  due to tidal effects is an open question
that requires a larger sample of systems before a firm conclusion may be drawn.
}
\end{figure}

In conclusion, we find that:  1) the assumption of an ``equilibrium tide" configuration
for the B-supergiant shape is a poor approximation;  2) the tidal effects produce orbital phase-dependent variations
in the line profiles leading to asymmetries such as extended blue or red wings; 3) the line-profile variability
causes the shape of the RV curve to depart significantly from that of a Keplerian RV curve, artificially 
enhancing the semi-amplitude; 4) the characteristics of the RV curve depend on the assumed rotation velocity 
of the B-supergiant.

Our analysis neglects the effects of a non-uniform temperature distribution over the stellar surface,  those
of non-radial pulsations and  the possible presence of enhanced mass-outflows after periastron passage.  These 
effects are expected to contribute even further towards deforming the line profiles.  Hence, we conclude that 
given these uncertainties and the results of our calculations, a low-mass neutron star ($m_{ns}\sim$1.5 M$_\odot$)  
cannot as yet be excluded for the Vela X-1 system.

\section{Acknowledgments}
The authors would like to acknowledge Matthieu Palate and the anonymous referee for helpful comments,
Alfredo Diaz and Ulises Amaya for computing assistance, and UNAM/PAPIIT project 107711-2 for financial
support.

\clearpage

\begin{table}[!h,!t,!b]
\begin{center}
\caption{Model input parameters and RV curve amplitudes. \label{tbl-2}}
\centering
\begin{tabular}{lllllllllllll}
\hline\hline
Name  &Sp.Type & $i$ & $\beta_0$ &    $R_1$ & $M_1$ & $\vv~sin~i$ & $e$ & $P$ & $M_{ns}$& References    \\
\hline
4U1538-52&B0Iab   & 68   &0.91$^{+0.20}_{-0.20}$         &15.7 &15.4 &180(20) & 0.17 & 3.73 &  0.874$\pm$0.073        & 4,5,6,7         \\
SMC X-1&  B0Ib    & 67   &0.91$^{+0.21}_{-0.19}$ &16.4 &15.7 &170(30) & 0.0  & 3.89 &  1.0$6^{+0.11}_{-0.10}$ & 1        \\
Her X-1& B--A,F   &$>$86 &1.00$^{+0.15}_{-0.15}$ & 3.8 & 2.0 &.....   & 0.0  & 1.70 &  1.07$\pm$0.358         & 7      \\
LMC X-4&  O8III   & 68   &0.97$^{+0.14}_{-0.12}$ & 7.8 &14.5 &240(25) & 0.0  & 1.41 &  1.25$^{+0.11}_{-0.10}$ & 1        \\
Cen X-3&O6.5II-III& 72   &0.75$^{+0.13}_{-0.13}$ &12.1 &20.2 &200(40) & 0.0  & 2.09 &  1.34$^{+0.16}_{-0.14}$ & 1         \\
Vela X-1& B0.5Iab & 79   &0.67$^{+0.04}_{-0.04}$ &31.8 &24.0 &116(6)  & 0.09 & 8.96 &  1.77$\pm$0.083         & 4,7,8,9    \\
Cyg X-2 &A9III    & 62   &1.2$^{+0.1}_{-0.1}$  & 7.0 & 0.6 & 34(2.5)& 0.0: & 9.84 &  1.78$\pm$0.23          & 10, 11     \\
\hline
\hline
\end{tabular}
\end{center}
References: $^1$van der Meer et al. (2007) and references therein; $^4$Rawls et al. 2011; $^5$Clark 2000;
$^6$Mukherjee et al. 2006; $^7$van Kerkwijk et al. 1995;  $^8$Barziv et al. 2001; $^9$Kreykenbohm et al. 2008;
$^{10}$Podsiadlowski\& Rappaport 2000; $^{11}$Casares et al. 1998.
\end{table}


\end{document}